\documentclass[journal]{IEEEtran}
\usepackage[numbers,sort&compress]{natbib}
\usepackage{bm}
\usepackage{url}
\usepackage{stfloats}
\usepackage{array}
\usepackage{graphicx}
\usepackage{epstopdf}
\usepackage{setspace}
\usepackage{lettrine}
\usepackage{amsmath}
\usepackage{amsbsy}
\usepackage{mathrsfs}
\usepackage{amsfonts,amssymb}
\usepackage{algorithm}
\usepackage{algorithmic}
\usepackage{graphicx}
\usepackage{subfigure}
\usepackage{booktabs}
\usepackage{amsmath}
\usepackage{amssymb}

\usepackage[justification=centering]{caption}

\usepackage{balance}
\usepackage{color}

\begin{document}

\title{\huge {Asymmetric Correntropy for Robust Adaptive Filtering}}

\author{Badong Chen,~\IEEEmembership{Senior Member,~IEEE,}
        Yuqing Xie,
        Zhuang Li,~\IEEEmembership{Student Member,~IEEE,}\\
        Yingsong Li,~\IEEEmembership{Senior Member,~IEEE,}
        Pengju Ren,~\IEEEmembership{Member,~IEEE}
\thanks{This work was supported by National Natural Science Foundation of China (91648208, 61976175), National Natural Science Foundation-Shenzhen Joint Research Program (U1613219), and The Key Project of Natural Science Basic Research Plan in Shaanxi Province of China (2019JZ-05).}
\thanks{Badong Chen, Yuqing Xie, Zhuang Li and  Pengju Ren are with the Institute of Artificial Intelligence and Robotics, Xi'an Jiaotong University, Xi'an, China. (chenbd@mail.xjtu.edu.cn, felixxyq@stu.xjtu.edu.cn, lizhuang771@stu.xjtu.edu.cn, pengjuren@mail.xjtu.edu.cn). Zhuang Li is also with the Xi'an Microelectronic Technology Institute, Xi'an, China.}
\thanks{Yingsong Li (liyingsong@ieee.org) is with the College of Information and Communication Engineering, Harbin Engineering University, Harbin 150001, China and also with the National Space Science Center, Chinese Academy of Sciences, Beijing 100190, China.}
}

\maketitle

\begin{abstract}
In recent years, correntropy has been seccessfully applied to robust adaptive filtering to eliminate adverse effects of impulsive noises or outliers. Correntropy is generally defined as the expectation of a Gaussian kernel between two random variables. This definition is reasonable when the error between the two random variables is symmetrically distributed around zero. For the case of asymmetric error distribution, the symmetric Gaussian kernel is however inappropriate and cannot adapt to the error distribution well. To address this problem, in this brief we propose a new variant of correntropy, named asymmetric correntropy, which uses an asymmetric Gaussian model as the kernel function. In addition, a robust adaptive filtering algorithm based on asymmetric correntropy is developed and its steady-state convergence performance is analyzed. Simulations are provided to confirm the theoretical results and good performance of the proposed algorithm.
\end{abstract}


\IEEEpeerreviewmaketitle

\section{Introduction}
\IEEEPARstart{A}daptive filters are used in a wide range of signal processing including channel equalization, noise cancellation, system modeling and so on. In recent years, much attention has been paid to the problem of how to improve the robustness of adaptive filters against impulsive noises \cite{chen2016generalized, lu2018performance, chambers1997robust, zayyani2014continuous, liu2007correntropy, chen2015convergence, shi2014convex, chen2017maximum, singh2009using, ma2015maximum}. To deal with impulsive noises, some robust adaptive filters were developed based on $p$-norm or mixed-norm, such as least mean $p$-norm (LMP) \cite{lu2018performance}, robust mixed-norm (RMN) \cite{chambers1997robust} and continuous mixed $p$-norm \cite{zayyani2014continuous}. Another efficient approach to solve this problem is to apply some exponential loss functions \cite{dennis1978techniques}, such as the Gaussian function in correntropy, to construct new cost functions for adaptive filters \cite{liu2007correntropy,chen2015convergence,shi2014convex,chen2017maximum,singh2009using,ma2015maximum}. Correntropy is a local similarity measure between two random variables defined in kernel space, which is insensitive to large errors and thus can suppress the adverse effects of impulsive noises or outliers \cite{principe2010information}.

The Gaussian function is generally adopted as the kernel function in correntropy, which is smooth and symmetric and has many desirable properties. In the information theoretic learning \cite{principe2010information}, the idea of probability density function (PDF) matching has been explored with great success. That is, the optimal kernel parameters can be chosen such that the kernel function is as close as possible to the error’s PDF. For the case of symmetric error distribution, the symmetric Gaussian kernel function can easily approximate the PDF of the error, and the original correntropy usually performs well. However, when the system is disturbed by noises of asymmetric distribution, which is widespread in many applications such as wireless localization \cite{youn2019robust}, the error distribution is usually asymmetric, and the symmetric Gaussian kernel is difficult to approximate the PDF of error. In this situation, the performance of the original correntropy may degrade seriously. To address this issue, we propose in this brief a new variant of correntropy, called asymmetric correntropy, which uses an asymmetric Gaussian model as the kernel function. The asymmetrically distributed error can be viewed as consisting of positive and negative error. The asymmetric Gaussian model has two kernel width parameters, which can approximate the PDFs of positive and negative parts respectively. Hence, compared with the symmetric Gaussian model, the asymmetric Gaussian model can get closer to the PDF of error, which suggests that the asymmetric Gaussian model can achieve performance improvement especially under the asymmetric noise. Based on the proposed asymmetric correntropy, a new robust adaptive filtering algorithm is developed and its steady-state convergence performance is analyzed.

The rest of the brief is organized as follows. In section II, we define the asymmetric correntropy and describe the maximum asymmetric correntropy criterion (MACC). In section III, we derive the adaptive filtering algorithm under MACC and analyze its steady-state convergence performance. Simulations are provided in section IV and conclusion is given in section V.

\section{ASYMMETRIC CORRENTROPY}
\subsection{Definition}
Given two random variables $X \in \mathbb{R}$ and $Y \in \mathbb{R}$ with joint PDF $ p_{XY}(x,y) $, correntropy is defined by \cite{chen2016generalized}
\begin{equation}\label{Eq1}
{V(X,Y)=E\left[\kappa (X,Y)\right]=\int\!\!\int\kappa (x,y) p_{XY}(x,y) dxdy},
\end{equation}
where $ \kappa (.,.) $ is usually a radial kernel, and $ E[.] $ denotes the expectation operator. Without explicit mention, the kernel function in correntropy is the well-known symmetric Gaussian kernel:
\begin{equation}\label{Eq2}
\kappa (X,Y)=G_\sigma (e)= exp\left( -\frac{e^2}{2 \sigma^2}\right),
\end{equation}
where $ e=X -Y $ is the error between $ X $ and $ Y $, and $ \sigma $ is the kernel bandwidth ($\sigma> 0$). In \cite{kato2002asymmetric}, an asymmetric Gaussian model was proposed, which can capture spatially asymmetric distributions. Inspired by the asymmetric Gaussian model, we propose in this work the concept of asymmetric correntropy, which adopts the following asymmetric Gaussian model as the kernel function:\\
\begin{equation}\label{Eq3}
\kappa(X,Y)=A_{\sigma _+ \sigma_-}(e)=\left\{
 \begin{aligned}
 &exp\left( -\frac{e^2}{2 \sigma_+^2}\right) ,  \quad if\ e  \ge 0\\
 &exp\left( -\frac{e^2}{2 \sigma_-^2}\right) ,  \quad if\ e  < 0
 \end{aligned}
 \right.
 \end{equation}
where $ \sigma_+ $ and $ \sigma_- $ denote, respectively, the bandwidths corresponding to the positive and negative parts of the error variable. In this study, we denote the asymmetric correntropy as
\begin{equation}\label{Eq4}
V_A (X,Y)=E\left[A_{\sigma _+ \sigma_-}(e)\right].
\end{equation}

\textit{Remark}: The kernel function in (3) is asymmetric and hence is not a Mercer kernel. As stated in \cite{chen2016generalized}, the kernel function in correntropy is not necessarily a Mercer kernel.

A desirable feature of the asymmetric Gaussian function in (3) is that it maintains the continuity and differentiability. Fig.1 shows the curves of the symmetric Gaussian function ($ \sigma=1.0 $) and asymmetric Gaussian function ($ \sigma_+=0.5, \sigma_-=2.0$). Clearly, when $ \sigma_+=\sigma_-$, the asymmetric Gaussian function will become the symmetric Gaussian function, and in this case the asymmetric correntropy will reduce to the original correntropy. In addition, it is easy to see that the asymmetric correntropy $ V_A (X,Y) $ is positive and bounded, namely $ 0<V_A (X,Y) \le 1$, which reaches its maximum if and only if $ X = Y $.
\begin{figure}[htb]
	\setlength{\abovecaptionskip}{0pt}
	\setlength{\belowcaptionskip}{0pt}
	\centering
	\includegraphics[height=1.8in]{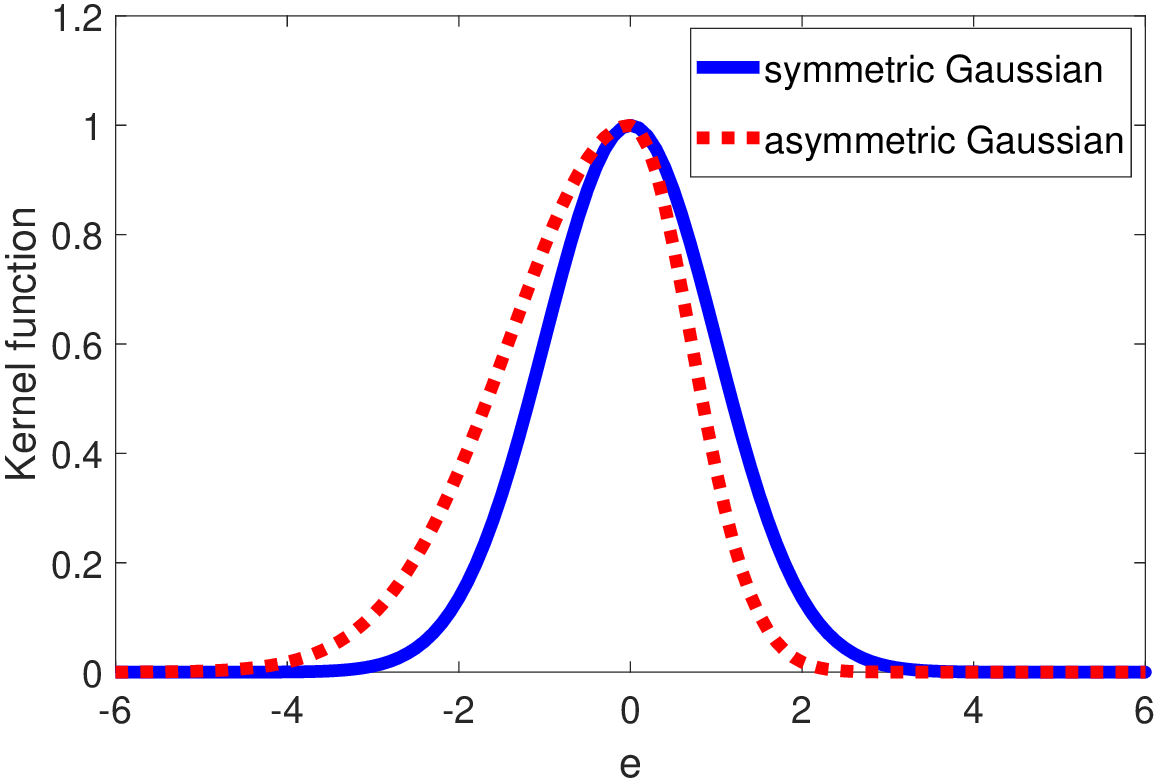}
	\caption{Curves of symmetric Gaussian function ($ \sigma=1.0 $) and asymmetric Gaussian function ($ \sigma_+=0.5, \sigma_-=2.0$)}
	\label{fig1}
\end{figure}

\subsection{Maximum asymmetric correntropy criterion}
Similar to the original correntropy, the proposed asymmetric correntropy can also be used as a robust cost function for various signal processing and machine learning tasks. For example, in supervised learning, the model can be trained to maximize the asymmetric correntropy between the model output and the target signal. We call this learning strategy the maximum asymmetric correntropy criterion (MACC). Let's now consider a supervised learning setting where the goal is to estimate the weight vector of a simple linear regression model:
\begin{equation}\label{Eq5}
y=\bm{\beta}^T \bm{x},
\end{equation}
in which $ \bm{\beta} \in \mathbb{R}^m $ is an $m$-dimensional weight vector, $ \bm{x} \in \mathbb{R}^m $  is the input vector and $ y \in \mathbb{R} $ is the model output. Given $N$ input-target training samples $\left\{\bm{x}_i, d_i\right\}^N_{i=1}$, the learning problem under MACC can thus be formulated by
\begin{equation}\label{Eq6}
\begin{aligned}
\bm{\beta}=&\mathop{arg\ max} \limits_{\bm{\beta} \in \mathbb{R}^m }  \hat{V}_A (y,d),\\
\hat{V}_A (y,d)&=\frac{1}{N} \sum_{i=1}^{N} A_{\sigma _+ \sigma_-}(e_i),
\end{aligned}
\end{equation}
where $ \hat{V}_A (y,d) $ is the estimated asymmetric correntropy between model output $y$ and target (desired) signal $d$, and $ e_i=d_i-y_i=d_i-\bm{\beta}^T \bm{x}_i $ is the $i$-th error sample. In practice, one can add a regularization term to the cost function and obtain
\begin{equation}\label{Eq7}
\begin{aligned}
&\bm{\beta}=\mathop{arg\ max} \limits_{\bm{\beta} \in \mathbb{R}^m } J(\bm{\beta}),\\
J(\bm{\beta})=& \frac{1}{N} \sum_{i=1}^{N} A_{\sigma _+ \sigma_-}(e_i) -\lambda ||\bm{\beta}||^2,
\end{aligned}
\end{equation}
where $\lambda \ge 0$ is the regularization parameter, and $||.||$ stands for Euclideam norm. Setting $ \frac{\partial{J(\bm{\beta})}}{\partial \bm{\beta}}=0$, we have

\begin{equation}\label{Eq8}
	\begin{aligned}
		&\frac{1}{N} \sum_{i=1}^{N} \frac{\partial}{\partial e_i} A_{\sigma _+ \sigma_-}(e_i) \frac{\partial e_i}{\partial \bm{\beta}} -2\lambda \bm{\beta} =0\\
		\Rightarrow &\frac{1}{N} \sum_{i=1}^{N} \frac{\partial}{\partial e_i} A_{\sigma _+ \sigma_-}(e_i) \bm{x}_i +2\lambda \bm{\beta} =0\\
		\Rightarrow &\frac{1}{N} \sum_{i=1}^{N} -\xi_{\sigma _+ \sigma_-}(e_i) e_i \bm{x}_i +\lambda \bm{\beta} =0\\
		\Rightarrow &\frac{1}{N} \sum_{i=1}^{N} \xi_{\sigma _+ \sigma_-}(e_i) (\bm{\beta}^T \bm{x}_i-d_i) \bm{x}_i +\lambda \bm{\beta} =0\\
		\Rightarrow & (A+\lambda I)\bm{\beta}=B\\
		\Rightarrow & \bm{\beta}=(A+\lambda I)^{-1} B
	\end{aligned}
\end{equation}
where $A=\frac{1}{N} \sum\limits_{i=1}^{N} \xi_{\sigma _+ \sigma_-}(e_i) \bm{x}_i x^T_i$, $B=\frac{1}{N} \sum\limits_{i=1}^{N} \xi_{\sigma _+ \sigma_-}(e_i) d_i \bm{x}_i$, and
\begin{equation}\label{Eq9}
\xi _{\sigma _+ \sigma_-}(e_i)=\left\{
\begin{aligned}
&\frac{1}{2 \sigma_+^2}exp\left(-\frac{e_i^2}{2 \sigma_+^2}\right) ,  \quad if\ e_i  \ge 0\\
&\frac{1}{2 \sigma_-^2}exp\left(-\frac{e_i^2}{2 \sigma_-^2}\right) ,  \quad if\ e_i  < 0
\end{aligned}
\right.
\end{equation}

\textit{Remark}: Similar to the optimal solution under the maximum correntropy criterion (MCC), the optimal solution under MACC is also a fixed-point solution of the equation $ \bm{\beta}=(A+\lambda I)^{-1} B $. Of course, this solution can be searched by a fixed-point iterative algorithm \cite{chen2015convergence}.

\begin{figure*}[b]
	\hrulefill
	\begin{align*}\label{cost11}
	S\approx \frac{\mu Tr(R_{x}) E\left[\psi _{\sigma _+ \sigma_-}^2(v)\right]}{2E\left[\psi _{\sigma _+ \sigma_-}^\prime(v)\right] -\mu Tr(R_{x})E\left[\psi _{\sigma _+ \sigma_-}(v) \psi _{\sigma _+ \sigma_-}^{\prime\prime}(v) +\left|\psi _{\sigma _+ \sigma_-}^\prime(v) \right|^2\right]}
	\tag{16}
    \end{align*}
\end{figure*}

\section{ADAPTIVE FILTERING UNDER MACC}
\subsection{MACC algorithm}
An alternative approach to search the optimal solution of the weight vector $\bm{\beta}$ under MACC is to use a stochastic gradient based adaptive filtering algorithm, which is computationally much simpler than the fixed-point algorithm and can track time-varying systems due  to  its  online  character. Based on the instantaneous cost function $\hat{J}(\bm{\beta})=A_{\sigma _+ \sigma_-}(e_i) $, a simple stochastic gradient algorithm, called in this brief the MACC algorithm, can easily be derived as follows:
\begin{equation}\label{Eq10}
\begin{aligned}
\bm{\beta}_{i+1}&=\bm{\beta}_i+ \mu \frac{\partial}{\partial \bm{\beta}_i} A_{\sigma _+ \sigma_-}(e_i)\\
&=\bm{\beta}_i+ \mu \ \psi_{\sigma _+ \sigma_-}(e_i) \bm{x}_i,
\end{aligned}
\end{equation}
where $\bm{\beta}_i$ denotes the estimated weight vector at the $i$-th iteration, $ e_i=d_i-\bm{\beta}^T \bm{x}_i$ is the prediction error, $\mu > 0$ is the step-size parameter, and the function $\psi_{\sigma _+ \sigma_-}(e_i) $ is
\begin{equation}\label{Eq11}
\psi _{\sigma _+ \sigma_-}(e_i)=\left\{
	\begin{aligned}
		&\frac{e_i}{\sigma_+^2}exp\left(-\frac{e_i^2}{2 \sigma_+^2}\right) ,  \quad if\ e_i  \ge 0\\
		&\frac{e_i}{\sigma_-^2}exp\left(-\frac{e_i^2}{2 \sigma_-^2}\right) ,  \quad if\ e_i  < 0
	\end{aligned}
\right.
\end{equation}

\textit{Remark}: When $\sigma_+=\sigma_- $, the derived MACC algorithm will become the MCC algorithm proposed in \cite{singh2009using}. In addition, the MACC algorithm can be viewed as the least mean square (MMSE) algorithm with variable step-size:
\begin{equation}\label{Eq12}
\mu_i =\left\{
\begin{aligned}
&\frac{\mu}{\sigma_+^2}exp\left(-\frac{e_i^2}{2 \sigma_+^2}\right) ,  \quad if\ e_i  \ge 0\\
&\frac{\mu}{\sigma_-^2}exp\left(-\frac{e_i^2}{2 \sigma_-^2}\right) ,  \quad if\ e_i  < 0
\end{aligned}
\right.
\end{equation}

\subsection{Steady-state convergence performance}

Now we analyze the steady-state convergence performance of the MACC algorithm. Suppose the target signal is generated by
\begin{equation}\label{Eq13}
d_i=\bm{\beta}^{*T}\bm{x}_i+v_i,
\end{equation}
where $ \bm{\beta}^{*} $ denotes an unknown weight vector that needs to be estimated, and $ v_i $ stands for a disturbance noise. In this case, we have $ e_i={e_a}(i)+v_i $, where ${e_a}(i)$ is the so-called \textit{a priori} error \cite{chen2014steady}. The mean-square \textit{a priori} error $E[e^2 _a(i)]$ is called the    (EMSE), which is a popular performance index of adaptive filters. The steady-state EMSE is $S={\lim\limits_{i\to \infty}} E[e^2 _a(i)]$. According to \cite{chen2016generalized}, if the step-size $\mu$ is chosen such that for all $i$
\begin{equation}
  \mu \leq \frac{2 E[e_{a}(i) \psi_{\sigma _+ \sigma_-}(e_i)]}{E[\|\bm{x}_i\|^{2} \psi_{\sigma _+ \sigma_-}^{2}(e_i)]},
\end{equation}
then the sequence of weight error power ($\text{WEP}=E[\|\bm{\beta}^{*}-\bm{\beta}_i\|^{2}]$) will be decreasing and converging.

At steady-state, the distributions of $e_a(i)$ and $e_i$ are both independent of $i$, we can omit the time index $i$ for brevity.
\iffalse
Therefore, (\ref{Eq11}) can be rewritten as
\begin{equation}\label{Eq14}
\psi _{\sigma _+ \sigma_-}(e)=\left\{
\begin{aligned}
&\frac{e}{\sigma_+^2}exp\left(-\frac{e^2}{2 \sigma_+^2}\right) ,  \quad if\ e  \ge 0\\
&\frac{e}{\sigma_-^2}exp\left(-\frac{e^2}{2 \sigma_-^2}\right) ,  \quad if\ e  < 0
\end{aligned}
\right.
\end{equation}
\fi
Taking the Taylor expansion of $\psi _{\sigma _+ \sigma_-}(e)$ with respect to $e_a$ around $v$ yields
\begin{equation}\label{cost10}
\begin{aligned}
\!\!\!\!\!&\psi _{\sigma _+ \sigma_-}(e) = \psi _{\sigma _+ \sigma_-}(e_a +v)\\
\!\!\!\!\!&=\psi _{\sigma _+ \sigma_-}(v)+\psi _{\sigma _+ \sigma_-}^\prime(v)e_a +\frac{1}{2} \psi _{\sigma _+ \sigma_-}^{\prime\prime}(v) e^2_a +o(e^2_a),
\end{aligned}
\end{equation}
\noindent where $\psi _{\sigma _+ \sigma_-}^\prime(v)$ and $\psi _{\sigma _+ \sigma_-}^{\prime\prime}(v)$ are the first and the second derivatives of $\psi _{\sigma _+ \sigma_-}(v)$, and $o(e^2_a)$ denotes the third and higher-order terms. Assume that  $e_a(i)$ is relatively small at steady-state and the third and higher-order terms in (\ref{cost10}) are negligible. Then we can obtain an approximate value of $S$, shown at the bottom of the page, where $Tr(.)$ is the trace operator and $R_{x}=E\left[\bm{x}_i  \bm{x}_i^T\right]$ is the covariance matrix of the input vector. For a detailed derivation, see literature \cite{chen2014steady}. The derivatives $\psi _{\sigma _+ \sigma_-}^\prime(v)$ and $\psi _{\sigma _+ \sigma_-}^{\prime\prime}(v)$ are

\begin{equation}\label{cost13}
\setcounter{equation}{17}
\begin{aligned}
\!\!\!\!\!\psi _{\sigma _+ \sigma_-}^{\prime}(v)=\!\!\left\{
				\begin{aligned}
				&\frac{1}{\sigma_+^2}exp\left(-\frac{{v^2}}{2\sigma_+^2}\right)\!\!\!\left(1-\frac{v^2}{\sigma_+^2}\right),\  if\ v  \ge 0\\
				&\frac{1}{\sigma_-^2}exp\left(-\frac{{v^2}}{2\sigma_-^2}\right)\!\!\!\left(1-\frac{v^2}{\sigma_-^2}\right),\  if\ v  < 0
				\end{aligned}
				\right.
\end{aligned}
\end{equation}
\begin{equation}\label{cost14}
\begin{aligned}
\!\!\!\!\!\!\psi _{\sigma _+ \sigma_-}^{\prime\prime}(v)=\!\!\left\{
						\begin{aligned}
						\!\!&\frac{1}{\sigma_+^2}exp\left(\!\!-\frac{{v^2}}{2\sigma_+^2}\right)\!\!\!\left(\frac{v^3}{\sigma_+^4}-\frac{3v}{\sigma_+^2}\right),\  if\ v  \ge 0\\
						\!\!&\frac{1}{\sigma_-^2}exp\left(\!\!-\frac{{v^2}}{2\sigma_-^2}\right)\!\!\!\left(\frac{v^3}{\sigma_-^4}-\frac{3v}{\sigma_-^2}\right),\  if\ v  < 0
						\end{aligned}
						\right.
\end{aligned}
\end{equation}

Given a noise distribution, the expectations $E\left[\psi _{\sigma _+ \sigma_-}^2(v) \right]$, $E\left[\psi _{\sigma _+ \sigma_-}^\prime(v)\right] $, and $E\left[\psi _{\sigma _+ \sigma_-}(v) \psi _{\sigma _+ \sigma_-}^{\prime\prime}(v) +\left|\psi _{\sigma _+ \sigma_-}^\prime(v) \right|^2\right] $ can be calculated by numerical integration. Then one can obtain theoretically an approximate value of S by using (\ref{cost11}).

\begin{figure}[tbp]
	\setlength{\abovecaptionskip}{0pt}
    \setlength{\belowcaptionskip}{0pt}
	\centering
	\includegraphics[height=1.8in]{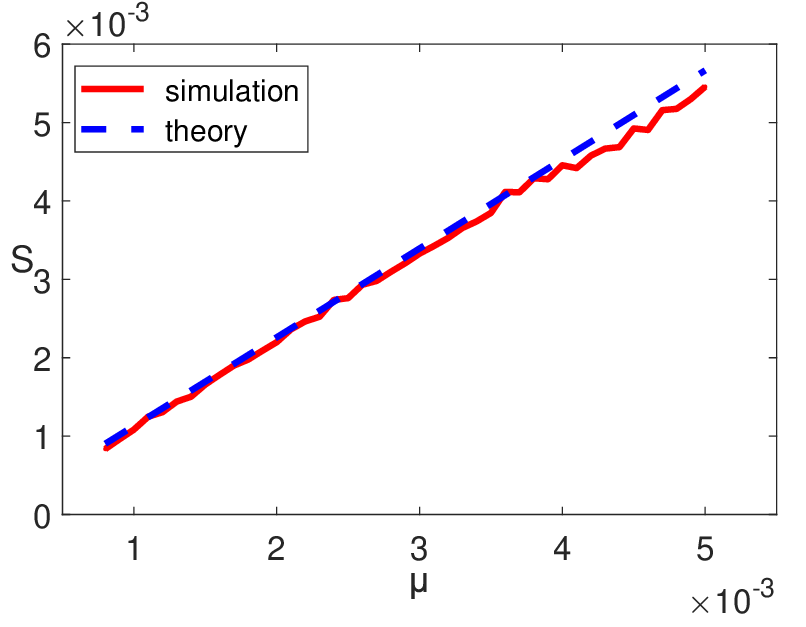}
	\caption{Theoretical and simulated EMSEs with different step-sizes}
	\label{fig2}
\end{figure}

\section{SIMULATIONS}
\subsection{Verification of theoretical results}
In the first subsection, we show the theoretical and simulated steady-state performance of the MACC algorithm. The weight vector of the unknown system is generated by a Gaussian distribution with mean 0.3 and variance 0.1. The initial weight vector of the adaptive filter is a null vector. The input signal is a zero-mean white Gaussian process with variance 1.0.

We consider a noise model with form $v_i=(1-a_i)A_i+a_i B_i$, where $a_i$ is a binary independent and identically distributed process with $Pr\{a_i=0\}=1-c$, with $0\leq c\leq1$ being an occurrence probability. $A_i$ is a process with small variance to represent the background noises, and $B_i$ is another process with much larger variance to represent the impulsive noises. $A_i$, $B_i$ and $a_i$ are mutually independent from each other. In the simulation, $c$ is set to $0.1$, the distributions of $A_i$ and $B_i$ are $\mathcal{N}(0,1.0)$ and $\mathcal{N}(0,400)$ respectively, where $\mathcal{N}(\mu,\sigma^2)$ denotes the Gaussian density function with mean $\mu$ and variance $\sigma^2$. The theoretical values (approximate values) of the steady-state EMSEs are calculated by using (\ref{cost11}), with different step-sizes. All the simulations are carried out on MATLAB 2016a running on the computer with i7-8700K, 3.70 GHZ CPU. Fig.\ref{fig2} shows the theoretical and simulated steady-state EMSEs with different step-sizes, where the simulated EMSEs are computed as an average over 100 independent Monte Carlo runs, and the steady-state value is obtained as an average over the last 500 iterations. To ensure the algorithm to reach the steady state, 40000 iterations are run in the simulation. One can observe: 1) the steady-state EMSEs are increasing with step-size; 2) when the step-size is small, the steady-state EMSEs computed by simulations match very well the theoretical values; 3) when the step-size become large, the experimental results will, however, gradually differ from the theoretical values, which coincides with the theoretical prediction (larger step-size causes larger error and makes the Taylor approximation poorer).

\begin{table*}[!t] \small
	\renewcommand\arraystretch{1.1}
	\setlength{\abovecaptionskip}{0pt}
	\setlength{\belowcaptionskip}{0pt}
	\centering
	\caption{Parameter settings of six algorithms}
	\begin{tabular}{cccccccccccccccc}
		\toprule $\quad$&SA\cite{mathews1987improved}&\multicolumn{4}{c}{LMM\cite{zou2000least}}&\multicolumn{2}{c}{LLAD\cite{sayin2014novel}}&\multicolumn{2}{c}{MCC\cite{singh2009using}}&\multicolumn{3}{c}{MCC-VC\cite{chen2019maximum}}&\multicolumn{3}{c}{MACC}\\
		\hline
		$\quad$&$\mu$&$\mu$&$\xi$&$\Delta_1$&$\Delta_2$&$\mu$&$\alpha$&$\mu$&$\sigma$&$\mu$&$\sigma$&$c$&$\mu$&$\sigma_-$&$\sigma_+$\\
		\hline
		Case 1)&0.005&0.01&0.5&6.0&6.2&0.007&1.8&0.033&1.15&0.033&1.15&0.35&0.025&2.6&0.8\\
		\hline
		Case 2)&0.0035&0.0085&0.4&8.0&10.2&0.0045&4.6&0.01&2.4&0.0078&3&0.56&0.023&3.0&1.2\\
		\bottomrule
	\end{tabular}
\end{table*}

\begin{figure*}[!t]
\setlength{\abovecaptionskip}{0pt}
\setlength{\belowcaptionskip}{0pt}
\centering
\subfigure[]{
\includegraphics[height=1.6in]{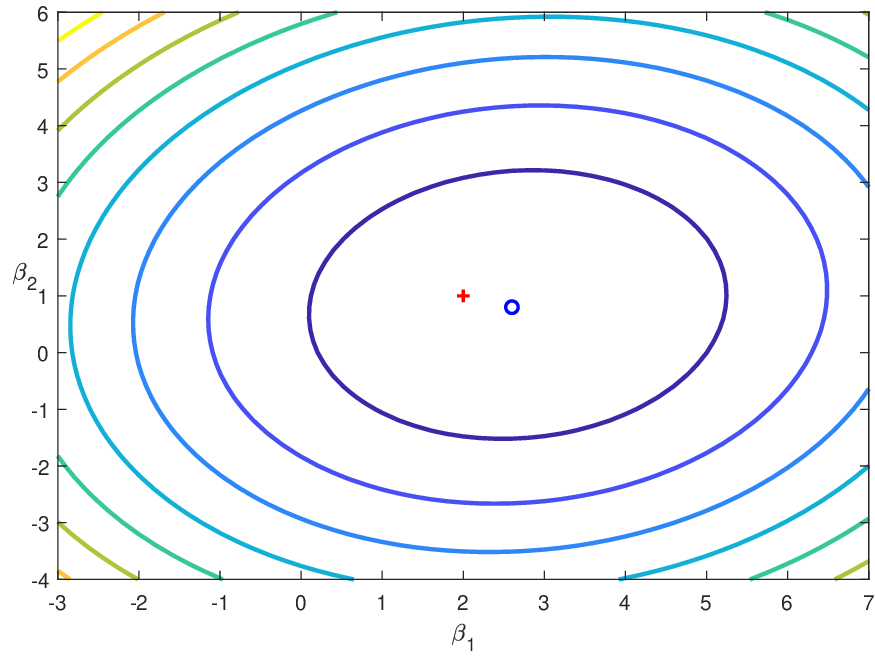}}
\subfigure[]{
\includegraphics[height=1.6in]{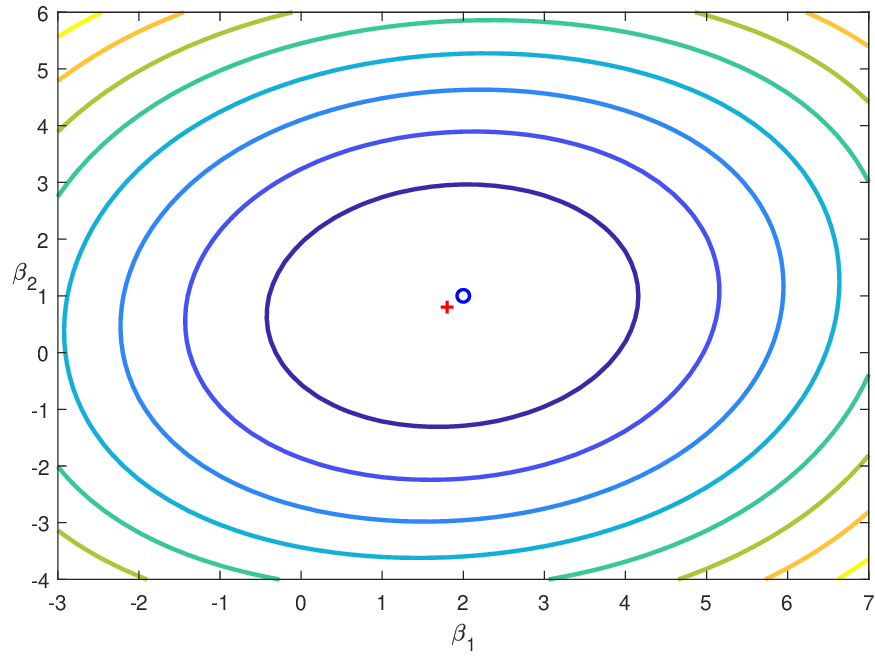}}
\subfigure[]{
\includegraphics[height=1.6in]{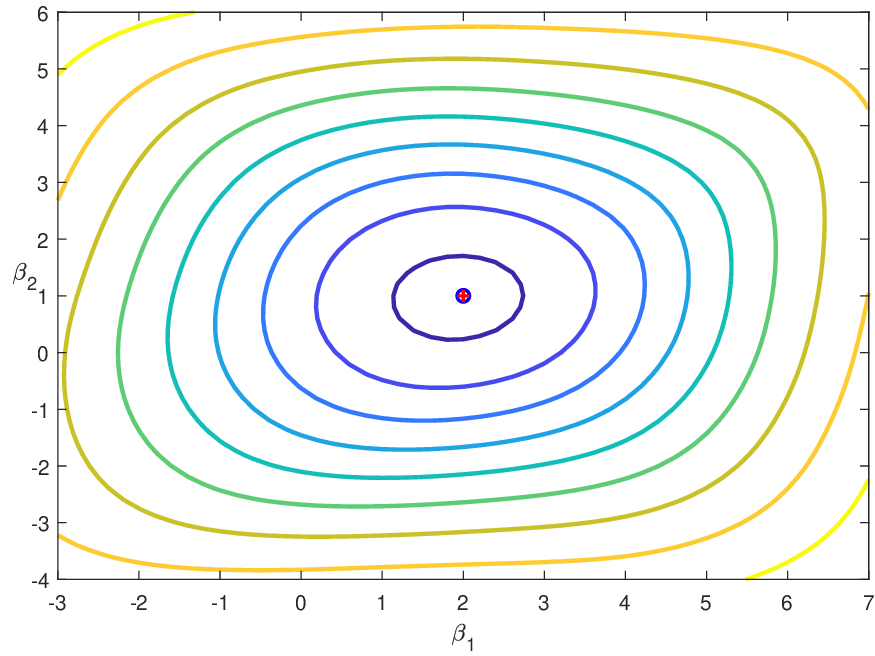}}
\caption{Contour plots of the performance surfaces. (a) MMSE, (b) MCC, (c) MACC.}
\label{fig3}
\end{figure*}

\begin{figure*}[!t]
	\setlength{\abovecaptionskip}{0pt}
	\setlength{\belowcaptionskip}{0pt}
	\centering
	\subfigure[]{
		\includegraphics[height=1.6in]{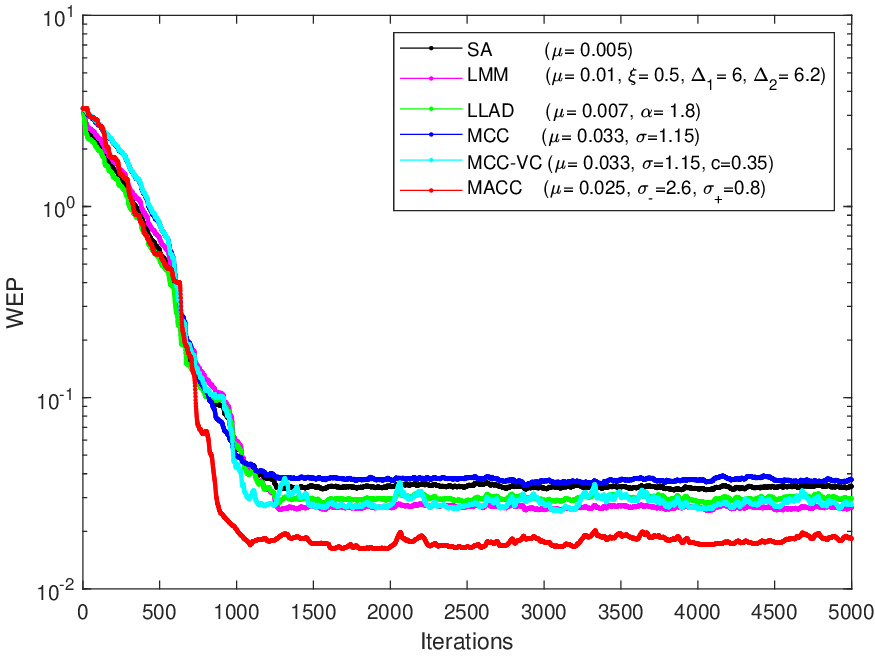}}
	\subfigure[]{
		\includegraphics[height=1.6in]{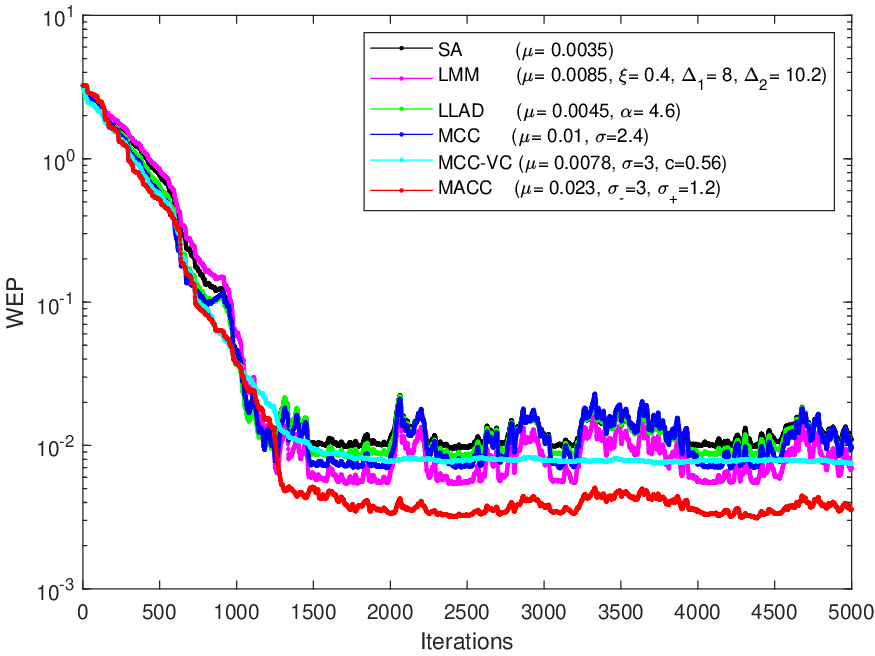}}
	\caption{Convergence curves in two cases: (a) asymmetric Gaussian density, (b) $F$ distribution.}
	\label{fig4}
\end{figure*}

\subsection{Performance comparison}
First, we show the contour plots of the performance surfaces for minimum mean squared error (MMSE), MCC \cite{singh2009using} and MACC in Fig.\ref{fig3} (i.e., the cost surfaces over the parameter space), in which the distribution of $A_i$ is assumed to be $F(5,8)$, where $F(n_1,n_2)$ denotes the $F$ distribution with degrees of freedom $n_1$ and $n_2$. Since the PDF of $F$ distribution is located on the positive half axis, in the simulation, we move the peak of the PDF to the origin. The red crosses in Fig.\ref{fig3} denote the target weight vector $\bm{\beta}^* =[2.0, 1.0]^T$, and the blue circles stand for the optimal weight vectors corresponding to the minima of the performance surfaces. As one can see, the optimal solutions under MMSE and MCC may deviate from the target value when faced with complex noises (e.g. asymmetric noises in this example). In particular, the MMSE solution can be far from the target value. However, the optimal solutions under MACC are almost the same as the target value, which implies that one can get a nearly unbiased estimate of the target weight vector under the MACC criterion.

Then, we compare the performance of the proposed MACC algorithm with several other representative robust adaptive filtering algorithms, including sign algorithm (SA) \cite{mathews1987improved}, LMM \cite{zou2000least}, LLAD \cite{sayin2014novel}, MCC and MCC-VC \cite{chen2019maximum}. For fairness, we select the different parameters for different algorithms so that all algorithms achieve their best performance with almost the same initial convergence speed. The parameter settings for two cases are given in Table I. Here, we consider two cases of the distribution of $A_i$: 1) asymmetric Gaussian density: $\mathcal{N} (0, 0.5) I (v<0) + \mathcal{N}(0, 5.0)I (v\ge 0) $, where $I$ stands for the indicator function; 2) $F$ distribution: $F(5,14)$. As before, the peak of the PDF for $F$ distribution is moved to the origin. The simulation results are obtained by averaging over 500 independent Monte Carlo runs. The performance measure adopted is the WEP. The convergence curves of the WEP are presented in Fig.\ref{fig4}. It is evident that the MACC algorithm can outperform (with much lower WEP) all other algorithms in both cases.

\section{CONCLUSION}
To deal with the data with asymmetric distributions, we proposed in this study a new variant of correntropy, called asymmetric correntropy, which uses an asymmetric Gaussian model as the kernel function. The supervised learning problem can thus be solved by maximizing the asymmetric correntropy between the model output and target signal. Under the maximum asymmetric correntropy criterion (MACC), a robust adaptive filtering algorithm, called MACC algorithm, was derived, and its steady-state convergence performance was analyzed. The analyzed results and desirable performance of the proposed algorithm were confirmed by simulations. An interesting topic for future study is how to determine the values of the free parameters $\sigma_+$ and $\sigma_-$ in practice.

\balance

\end{document}